\documentclass[pra,aps,longbibliography,showpacs,groupedaddress,superscriptaddress,twocolumn,toc=flat,nofootinbib]{revtex4-1}
\usepackage{graphicx}
\usepackage{latexsym}
\usepackage{amssymb}
\usepackage{amsmath}
\usepackage{amsfonts}
\usepackage{upgreek}
\usepackage{subfigure}
\usepackage{bm}
\usepackage{bbold}
\usepackage{verbatim}
\usepackage{siunitx}
\usepackage[unicode=true,
 bookmarks=false,
 breaklinks=false,pdfborder={0 0 1},backref=false,colorlinks=true]
 {hyperref}
\hypersetup{
 linkcolor=[rgb]{0,0,1},citecolor=[rgb]{0,0,1},urlcolor=[rgb]{0,0,1}}
\usepackage{multirow}
\usepackage{color}
\usepackage{comment}
\usepackage{xcolor}
\usepackage{soul}
\usepackage{tikz}
\usepackage{tabularx}
\usepackage{txfonts}
\allowdisplaybreaks

\renewcommand{\H}{\hat{H}}

\usepackage{braket}

\begin{document}

\title{Supersolidity in Rydberg tweezer arrays}

\author{Lukas~Homeier}
\email{lukas.homeier@jila.colorado.edu}
\affiliation{Department of Physics and Arnold Sommerfeld Center for Theoretical Physics (ASC), Ludwig-Maximilians-Universit\"at M\"unchen, Theresienstr. 37, M\"unchen D-80333, Germany}
\affiliation{Munich Center for Quantum Science and Technology (MCQST), Schellingstr. 4, M\"unchen D-80799, Germany}
%\affiliation{JILA and Department of Physics, University of Colorado, Boulder, CO, 80309, USA}
%\affiliation{Center for Theory of Quantum Matter, University of Colorado, Boulder, CO, 80309, USA}

\author{Simon~Hollerith}
\affiliation{Department of Physics, Harvard University, Cambridge, Massachusetts 02138, USA}

\author{Sebastian~Geier}
\affiliation{Physikalisches Institut, Universit\"at Heidelberg, Im Neuenheimer Feld 226, 69120 Heidelberg, Germany}

\author{Neng-Chun~Chiu}
\affiliation{Department of Physics, Harvard University, Cambridge, Massachusetts 02138, USA}

\author{Antoine~Browaeys}
\affiliation{Universit{\'e} Paris-Saclay, Institut d’Optique Graduate School, CNRS, Laboratoire Charles Fabry, 91127 Palaiseau Cedex, France}

\author{Lode~Pollet}
\email{lode.pollet@physik.uni-muenchen.de}
\affiliation{Department of Physics and Arnold Sommerfeld Center for Theoretical Physics (ASC), Ludwig-Maximilians-Universit\"at M\"unchen, Theresienstr. 37, M\"unchen D-80333, Germany}
\affiliation{Munich Center for Quantum Science and Technology (MCQST), Schellingstr. 4, M\"unchen D-80799, Germany}

\date{\today}
\begin{abstract}
%Strong interactions in quantum simulators enable the realization and study of quantum matter. 
Rydberg tweezer arrays provide a versatile platform to explore quantum magnets with dipolar XY or van-der-Waals Ising ZZ interactions. Here, we propose a scheme combining dipolar and van-der-Waals interactions between two Rydberg states, where the amplitude of the latter can be greater than that of the former, realizing an extended Hubbard model with long-range tunnelings in optical tweezer arrays.
On the triangular lattice with repulsive interactions, we predict the existence of a robust supersolid phase with a critical entropy per particle $S/N \approx 0.19$ accessible in current Rydberg tweezer experiments  supported by large-scale quantum Monte Carlo simulations. We further demonstrate the experimental feasibility by identifying pairs of Rydberg states in ${}^{87}$Rb realizing the required interactions. Such a lattice supersolid is long-lived, found over a wide parameter range in an isotropic and flat two-dimensional geometry, and can be realized for 100s of particles allowing one to directly probe the defect-induced picture of supersolids. Its thermodynamical and dynamical properties can hence be studied at a far larger scale than hitherto possible.
\end{abstract}
\maketitle

%%%%%%%%%%%%%%%%%%%%%%%%%%%%%%%%%%
\textbf{Introduction.---}
Supersolids are enigmatic states of matter in which the usually competing translational  and $U(1)$ symmetries are simultaneously broken~\cite{Boninsegni_RMP2012}. First conceived theoretically by E. Gross~\cite{Gross1957}, they can be pictured as a density modulated superfluid. Recent progress in cold atomic physics saw flashes of such supersolids~\cite{Chomaz2019,Tanzi2019,Boettcher2019,Pollet_newsandviews,Norcia2021} in elongated traps for a few droplets, following theoretical proposals~\cite{Wenzel2017,Baillie2018,Macia2016,Cinti2017, Chomaz2023}. Other realizations are based on cavity-mediated all-to-all interactions~\cite{Leonard2017,Landig2016,Klinder2015} or spin-orbit coupled Bose-Einstein condensates~\cite{Li2017}. There exists, however, a complementary view on supersolids, introduced by Andreev-Lifshitz-Chester~\cite{Andreev1969,Chester1970}, in which defects such as vacancies and interstitials may delocalize on top of a robust crystal with one atom per unit cell. As the original claims of bulk supersolidity in ${}^4$He are now understood differently~\cite{Kim2004_Nature,Kim2004,Kim2012}, the realization of supersolids with spontaneous symmetry breaking in the tight-binding limit has remained elusive.

Such lattice supersolids for hard-core bosons have a long theoretical history. Two decades ago, quantum Monte Carlo simulations~\cite{Boninsegni2003, Melko2005,Heidarian2005,Wessel2005,Boninsegni2005} confirmed mean-field predictions~\cite{Murthy1997} for the model with nearest-neighbor hopping and interactions, establishing the existence of a supersolid phase for any filling between 1/3 and 2/3 on the triangular lattice. In contrast, for the same model on the square lattice no supersolid phase is found, in line with the domain wall argument~\cite{Sengupta2005,Wessel2005}. 
%The supersolid phase can be seen as a quantum order-by-disorder scenario~\cite{Villain1980,Henley1989}.
The controversy at half filling was settled in favor of a first-order transition between the supersolids found above and below half filling~\cite{Heidarian2005,Boninsegni2005}. The model attracted renewed interest when polar molecules became a realistic option~\cite{Cornish2024}; models with dipolar repulsion found stable supersolids on the triangular~\cite{Pollet2010} and also on the square lattice~\cite{CapogrossoSansone2010}.
With recent experimental progress in magnetic atoms~\cite{Norcia2021,Su2023}, itinerant Rydberg dressing~\cite{Weckesser2024}, and polar molecules~\cite{Cornish2024}, testing these predictions will soon become reality, and offer an alternative to the spin supersolids concurrently sought in frustrated magnets with {\it anti-}ferromagnetic spin exchange~\cite{Sellmann2015,Gao2022,Tsurkan2017,Xiang2024,Chen2024_supersolid,Zhu2024_supersolid}.

Rydberg tweezer arrays are another experimental cold atom platform that has gained significant attention in both analog~\cite{Bernien2017,Scholl2021,Ebadi2021,Steinert2023,Chen2023,Shaw2024,Anand2024} and digital quantum simulation~\cite{Madjarov2020,Graham2022,Bluvstein2023}.
In the realm of analog simulation, studies have mainly focused on two classes of Hamiltonians~\cite{Browaeys2020}: (i) A pair of atoms in the same Rydberg state experiences a van-der-Waals (vdW) interaction that can be used to implement Ising ZZ spin models~$\propto J_z/r_{ij}^6$ \cite{Bernien2017,Scholl2021,Ebadi2021,Shaw2024,Anand2024}; here $r_{ij}$ is the distance between atoms at site~$i$ and~$j$. (ii) The direct dipole-dipole interaction between Rydberg states of different parity enables to study dipolar XY spin models~$\propto J_\perp/r_{ij}^3$ \cite{Chen2023,Bornet2023}; this model can be mapped to hard-core bosons~\cite{Leseleuc2019} with long-range tunnelings.
%Conveniently, this model can be mapped to hard-core bosons~\cite{Leseleuc2019} with long-range tunnelings~$t_{ij} = 2 J_\perp/r_{ij}^3$, but with no additional interactions besides the hard-core constraint.

\begin{figure*}[t!!]
\centering
\includegraphics[width=\linewidth]{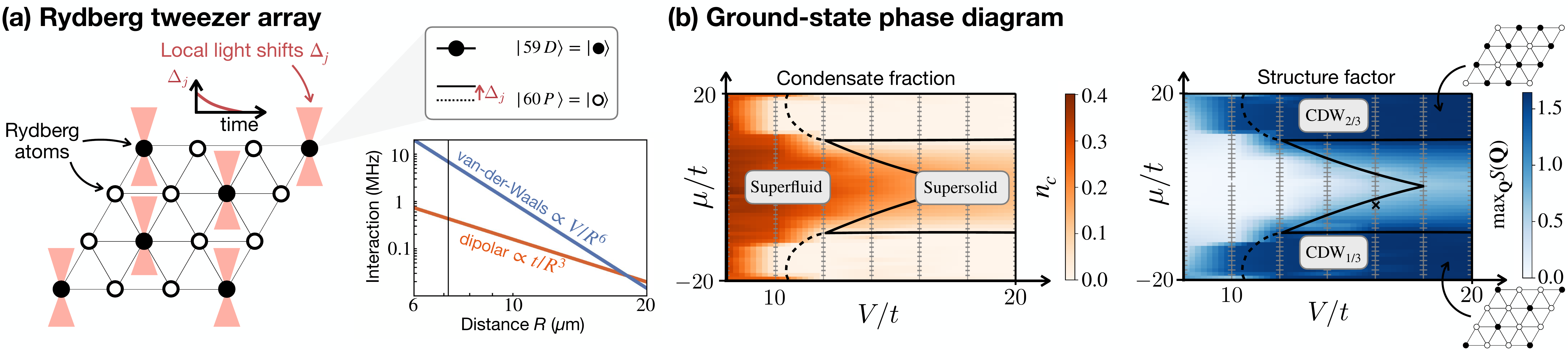}
\caption{\textbf{Dipolar and vdW model on the triangular lattice.} \textbf{(a)} We propose to implement an extended Hubbard model in a specific set of Rydberg states in ${}^{87}$Rb featuring strong vdW density-density interactions~$V/r_{ij}^{6}$ and competing dipolar tunneling amplitudes~$\propto -t/r_{ij}^{3}$. Equivalently, we can express the model as spin-1/2 particles with dipolar XY interactions and vdW Ising ZZ interactions. \textbf{(b)} Using quantum Monte Carlo simulations, we calculate the ground-state phase diagram for repulsive extended Hubbard interactions ($V>0$) and ferromagnetic tunneling amplitudes ($t>0$) on the triangular lattice; here, we tune the chemical potential~$\mu$ for particles in the system. We show the maximum structure factor $\mathrm{max}_{\bm{Q}} S(\bm{Q})$ (right) and the condensate fraction~$n_c$ (left) for a representative system size~$L_x \times L_y=15\times 15$ and at temperature~$\beta t = 5$. We find crystalline charge-density wave (CDW) order at 1/3 and 2/3 filling flanked by a supersolid region; the dashed (solid) black lines indicate first-order (second-order) phase transitions~\cite{Supplements}. The ratio~$V/t$ can be tuned by the lattice spacing~$R$. The cross indicates an experimentally feasible regime with~$R \approx 7.1\,\mu\mathrm{m}$ and~$t \approx 2\pi\times 0.45\,\mathrm{MHz}$ analyzed in Fig.~\ref{fig_temperature}. }
\label{fig-overview}
\end{figure*}

In this Letter, we propose a scheme enabling the implementation of hard-core bosons with competing dipolar tunnelings~$t$ and vdW extended Hubbard interactions~$V$ encoded in two Rydberg states~\cite{Hond2020,Zeybek2023}, combining the previously established models in tweezer arrays, see Fig.~\ref{fig-overview}(a).
For ferromagnetic tunnelings~$t>0$ and repulsive Hubbard interactions~$V>0$, we show the existence of a long-lived lattice supersolid on the triangular lattice using quantum Monte Carlo techniques, see Fig.~\ref{fig-overview}(b). 
Its critical temperature and entropy can be as high as $T_c/ t  \sim 2$ and ~$S_c/N \sim 0.19$, putting it well within experimental reach and without concerns about an elongated, confining geometry.
For its experimental realization, we propose a specific set of Rydberg states in ${}^{87}$Rb and an experimental protocol based on adiabatic state preparation techniques~\cite{Chen2023}, allowing one to prepare the supersolid in current Rydberg tweezer arrays with hundreds of atoms; hence to directly test the long-sought Andreev-Lifshitz-Chester picture in existing experimental technology. In this defect-induced picture of supersolidity, the energy gain through delocalizing and condensing vacancies is larger than their potential energy cost of distorting the ideal, commensurate solid, giving rise to simultaneous diagonal and off-diagonal long range order.

%%%%%%%%%%%%%%%%%%%%%%%%%%%%%%%%%%
\textbf{Model.---}
We consider hard-core bosons~$\hat{a}^\dagger_j$ with Hamiltonian
\begin{align}
\begin{split}
    \H &= \H_{t} + \H_{V} \label{eq:Hamiltonian-XY-ZZ} \\
    \H_{t} &= -t\sum_{i<j} \frac{1}{r_{ij}^3} \left( \hat{a}^\dagger_{i}\hat{a}_{j} +\mathrm{h.c.} \right) \\
    \H_{V} &= V\sum_{i<j} \frac{1}{r_{ij}^6} \left(\hat{n}_{i}-\frac{1}{2}\right)\left(\hat{n}_{j}-\frac{1}{2}\right) - \mu\sum_{j}  \hat{n}_j,
\end{split}
\end{align}
where~$\hat{n}_j = \hat{a}^\dagger_j\hat{a}_j$ is the number operator on lattice sites~$j$ of a triangular lattice, see Fig.~\ref{fig-overview}(a). The distance between two lattice sites is~$r_{ij} = R \cdot |i-j|$ and in the following we set the lattice constant~$R=1$.
Therefore, the Hamiltonian~\eqref{eq:Hamiltonian-XY-ZZ} describes hard-core bosonic matter with ferromagnetic, dipolar tunneling amplitudes~$t > 0$ and repulsive, extended vdW Hubbard interactions~$V > 0$.
The chemical potential~$\mu$ controls the filling of particles such that $\mu=0$ corresponds to half filling.

Let us first discuss the various ordered phases that may originate in the ground state, see Fig.~\ref{fig-overview}(b). For~$V=0$, we expect a condensate with true off-diagonal long-range order being possible in two dimensions even at finite temperature because of the dipolar nature of the tunneling amplitudes. The condensate is equivalent to an in-plane ferromagnet~\cite{Bernardet2002}, such as described by the XY model, and has been studied in Refs.~\cite{Chen2023,Sbierski2024}. For~$V/t = \infty$, the system orders in a crystalline, charge-density wave (CDW) with a wave ordering vector $\bm{Q}$ depending on the filling factor. In this work, we restrict ourselves to not too large values of $V/ t \lesssim 24$ where only the commensurate fillings 1/3 and 2/3 are relevant as shown in Fig.~\ref{fig-overview}(b) left.

When the dipolar tunneling and repulsive Hubbard interactions are in competition, we anticipate the existence of a supersolid as the commensurate structure with filling 1/3 is doped with particles (or, equivalently, as the CDW with filling 2/3 is doped with holes), in analogy to models with short-range exchange,~\cite{Boninsegni2003,Melko2005,Heidarian2005,Wessel2005,Boninsegni2005}, see Fig.~\ref{fig-overview}(b) left. Next, we corroborate the existence of the supersolid phase in numerical simulations.

%%%%%%%%%%%%%%%%%%%%%%%%%%%%%%%%%%
\textbf{Quantum Monte Carlo simulations.---}
We employ Quantum Monte Carlo (QMC) simulations with worm-type~\cite{Prokofev_worm_1998,Pollet2012} updates to study the model, Eq.~\eqref{eq:Hamiltonian-XY-ZZ}, with ferromagnetic tunnelings $t = 1$ on the triangular lattice of linear size $L$ with periodic boundary conditions (PBC) and inverse temperature~$\beta$.
The code is based on an adaptation of Ref.~\cite{Sadoune2022}. The simulations were found to be challenging because of the following reasons: (i) the parametric difference between the dipolar kinetic and potential vdW terms, (ii) strong first-order transitions between the superfluid (SF) and CDW phases, (iii) critical slowing down near the second-order crystalline transitions. This is reflected in the data where the shown statistical errors do not capture all the noise originating from the systematic errors.
%Some of the data shown took several CPU months before the simulations thermalized.
%
%In order to study the ${\rm U}(1)$ symmetry breaking we analyze the spin stiffness~$\chi_{\rm s}$, which is related to the fluctuations of the winding number as
%$\chi_{\rm s} = \left< W^2 \right> / (2 \beta)$~\cite{Pollock1987}.
In order to study the $U(1)$ symmetry breaking we analyze the experimentally accessible condensate fraction~$n_c$ defined by
\begin{equation}
n_c = \frac{1}{N} \sum_{d} C^{+-}(d),
\label{eq:def_condfrac}
\end{equation}
where $ C^{+-}(d) = \left< \hat{a}^\dagger_{j+d} \hat{a}_{j} + {\rm h.c.} \right>$ is the equal-time off-diagonal correlation function at distance~$d$ from an arbitrary reference site~$j$, which can be directly obtained in the proposed Rydberg tweezer setup by measuring the in-plane spin-spin correlations $\langle \hat{S}^x_i \hat{S}^x_j \rangle$ in the spin picture. The criticality of the analogous dipolar XY~model has been analyzed in Ref.~\cite{Sbierski2024}, from which we extract the critical exponents $\nu = \eta = 1$.

The lattice symmetry breaking transition to a $\sqrt{3}\times\!\sqrt{3}$ order is caught by the structure factor
\begin{equation} \label{eq:Sq}
    S(\bm{Q}) = \frac{1}{L^2}\left< \left| \sum_{k=1} \hat{n}_k e^{i \bm{Q} \bm{r}_k}  \right|^2 \right>,
\end{equation}
with ordering vector $\bm{Q} = (4\pi/3,0)$. Its critical exponent in a finite size scaling analysis is $2\beta/\nu$. The vdW interactions are short-ranged; hence, no changes to the known critical exponents are expected.

\begin{figure}[t!!]
\centering
\includegraphics[width=\linewidth]{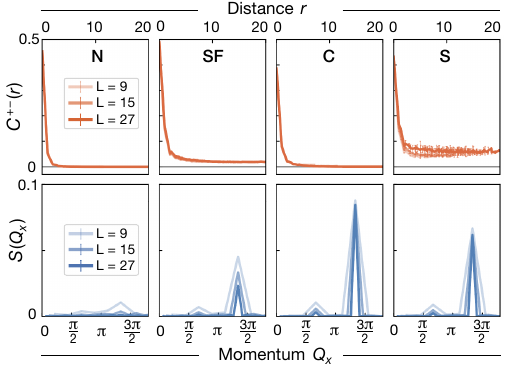}
\caption{\textbf{Correlation functions.} We show the off-diagonal correlation function $C^{+-}(r)$ (upper) and static structure factor $S(Q_x)$ (lower) in the four different phases: (N) normal ($\mu/t=-9, V/t=20, \beta t=0.1$), (SF) superfluid ($\mu/t=-1, V/t=20, \beta t=0.5$), (C) charge-density wave ($\mu/t=-9, V/t=20, \beta t=0.4$), and (S) supersolid ($\mu/t=-5, V/t=20, \beta t=0.6$). The system sizes are $L=9$, $15$ and $27$ (light to dark). Deep in the normal and CDW phases, $C^{+-}(r)$ decays asymptotically as $\sim r^{-3}$; in the superfluid and supersolid phases it is asymptotically constant. Only the peaks at $Q_x=4\pi/3$ in the static structure factor in the charge density wave and supersolid phases are indicative of crystalline order; the others shrink to zero.} 
\label{fig:illustr}
\end{figure}
%%%%%%%%%%%%%%%%%%%%%%%%%%%%%%%%%%
\textbf{Ground-state phase diagram.---} 
We perform simulations for linear system sizes~$L=9,15,27$ commensurate with the expected $1/3$ and~$2/3$-filling phases, at sufficiently low temperature~$\beta t=(T/t)^{-1}=10$, which probes the ground state of Hamiltonian~\eqref{eq:Hamiltonian-XY-ZZ} for all practical purposes. In Fig.~\ref{fig-overview}(b), we show the maximum structure factor~$\mathrm{max}_{\bm{Q}}S(\bm{Q})$ and the condensate fraction~$n_c$ for~$L=15$. In our simulations, we find CDW lobes at 1/3- and 2/3-filling. Finite size effects can be appreciated from the supplemental~\cite{Supplements}. In the ground state, the phase diagram bears a strong similarity to the models with a short-range exchange term~\cite{Boninsegni2003,Melko2005, Heidarian2005,Wessel2005,Boninsegni2005,Pollet2010}.

In the following, we focus on the vicinity of the ${\rm CDW}_{1/3}$ region, i.e., the lower lobe in Fig.~\ref{fig-overview}(b) right. In analogy to the defect picture of Andreev-Lifshitz-Chester, we start within the lobe of constant particle density and increase the chemical potential above $\mu/t \gtrsim 10$; hence doping the system with particles. For sufficiently large~$V/t \gtrsim 12$, we find a reduced but still substantial structure factor~$S(\bm{Q})$ for~$\bm{Q}=(4\pi/3,0)$; hence establishing long-range order in the density-density $\langle \hat{n}_i \hat{n}_j \rangle$ correlations. In addition, the dopants give rise to a condensate fraction~$n_c$; hence long-range order in the off-diagonal $\langle \hat{a}_i^\dagger \hat{a}_j +\mathrm{h.c.} \rangle$ correlations.
The co-existence of both order parameters confirms the existence of a supersolid phase in the ground state of our proposed model. 

The experimentally accessible static structure factor and the off-diagonal correlation function for the different phases are shown in Fig.~\ref{fig:illustr}, where we set~$V/t=20$ corresponding to a fixed distance between Rydberg atoms. Hence, we describe an experimentally realistic situation, where all the above described phases can be realized at finite temperature by a varying the chemical potential~$\mu$~\cite{Supplements}, and distinguished through their correlation functions.

We further characterize the phase transitions. In the vicinity of the tip of the lobe, the ${\rm CDW}_{1/3}$ shows a first-order phase transition directly into the SF phase without crossing a supersolid region, as indicated by the dashed line in Fig.~\ref{fig-overview}(b). As we increase~$V/t$ the first-order line splits into two lines of second-order transitions indicated by the solid lines in Fig.~\ref{fig-overview}(b): (i) The CDW to supersolid transition described above is the $U(1)$~symmetry breaking of the XY ferromagnet. (ii) At intermediate $V/t$, the supersolid loses the translational-symmetry broken crystalline structure as we approach~$\mu=0$ leading to a transition into the SF phase.

%For our representative point~$J_z=8$, a finite size scaling analysis ...
%In the Supplementary material, we provide a detailed analysis of the ground-state phase diagram ...

%The model with nearest-neighbor hopping and nearest-neighbor interaction attracted substantial interest two decades ago. Mean-field predictions~\cite{Murthy1997} were confirmed by quantum Monte Carlo simulations~\cite{Boninsegni2003, Melko2005, Heidarian2005, Wessel2005, Boninsegni2005} establishing the existence of a supersolid phase for any filling between 1/3 and 2/3 , in contrast to the same model on the square lattice where no supersolid phase is found, in line with the domain wall argument~\cite{Sengupta2005, Wessel2005}. The supersolid phase can be seen as a quantum order-by-disorder scenario~\cite{Villain1980,Henley1989}.
%The controversy at half filling was settled in favor of a first order transition between the supersolids found above and below half filling~\cite{Heidarian2005,Boninsegni2005}. The model attracted renewed interest when polar molecules became a realistic option; models with dipolar repulsion found stable supersolids on the triangular~\cite{Pollet2010} and also on the square lattice~\cite{CapogrossoSansone2010}.

\begin{figure}[t!!]
\centering
\includegraphics[width=\linewidth]{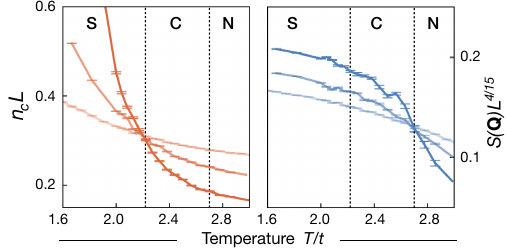}
\caption{\textbf{Finite temperature transitions.} We perform a finite size scaling analysis of the finite temperature transitions for $\mu/t = -4$, $V/t = 16$ from QMC simulations. The transition between the high temperature magnet ('N') and the insulating density wave phase ('C') belongs to the universality class of the 2D 3-state Potts model with $\nu = 5/6$ and $\beta = 1/9$ (right panel). The $U(1)$ transition to the supersolid phase ('S') has a mean-field exponent $\nu = 1$ (left panel). The system sizes are $L=9$, $15$ and $27$ (light to dark). Crossing points determine the respective critical temperatures; finite size effects are seen to be small. } 
\label{fig_temperature}
\end{figure}
%%%%%%%%%%%%%%%%%%%%%%%%%%%%%%%%%%
\textbf{Finite temperature transitions.---}
We choose a representative parameter point $\mu/t = -8, V/t = 16$ where we expect a supersolid phase. As shown in Fig.~\ref{fig_temperature}, translational symmetry is the first symmetry we break when lowering the temperature. The transition is well described by the (2+1D) three-state Potts model, with known critical exponents $\nu=5/6$ and $\beta=1/9$. When lowering the temperature further, a supersolid phase forms at $T_c/t=2.2$. 
A finite size analysis~\cite{Supplements} of the superfluid stiffness and of the condensate fraction yield the same critical temperature for the $U(1)$ transition, with the critical exponents are $\nu=\eta=1$, as expected from Refs.~\cite{Sbierski2024,Chen2023}.
The critical entropy per site was found to be $S/N \approx 0.14$ for $L=27$. For $V/t=20$, $\mu/t =-5$ the critical entropy was the highest within our parameter range and found to be $S/N \approx 0.19$ with a $T_c/t \approx 1.9$~\cite{Supplements}. These entropies are within reach for the current generation of Rydberg tweezer experiments~\cite{Chen2023,Sbierski2024, chen2023spectroscopy,Semeghini2021,zhenjiu2024}.

%%%%%%%%%%%%%%%%%%%%%%%%%%%%%%%%%%
\textbf{Experimental proposal.---}
%I think there is no need for the next sentence
%There exist several schemes how Rydberg interactions can be mapped to spin models (cite!). Here, 
We propose to encode the empty and occupied lattice sites in two Rydberg states of ${}^{87}$Rb.
The tunneling amplitudes $t = -C_3$ arise from resonant dipole-dipole interactions, while extended Hubbard interactions $V/R^6 =  E_{\medbullet \medbullet} + E_{\medcirc \medcirc} - 2E_{\medbullet \medcirc} = C^{\mathrm{eff}}_6/R^6$ emerge via second order van-der-Waals interactions~\cite{Whitlock_2017}. 
Here, $E_{\sigma \sigma'}\propto 1/R^6$ are the van-der-Waals interactions between a pair of atoms in state~$\ket{\,\sigma \sigma'}$ on neighbouring sites, and coefficients $C_3$ and $C^{\mathrm{eff}}_6$ define the interaction strength.  
%The Ising spin interaction is given by~$J_z =  E_{\uparrow \uparrow} + E_{\downarrow \downarrow} - 2E_{\uparrow \downarrow}$, where $E_{\sigma \sigma'} = C^{\sigma \sigma'}_6/a^6$ is the van-der-Waals interaction on neighbouring sites and $C^{\sigma \sigma'}_6$ describes the diagonal interaction between a pair of atoms in state~$\ket{\,\sigma \sigma'}$. The dipolar XY interaction is defined as~$J_\perp = 2C_3/a^6$ with direct dipole-dipole coupling~$C_3$.
The mapping from the Rydberg Hamiltonian to the extended Hubbard model introduces an additional position-dependent chemical potential~$\mu_j = \frac{1}{2}\sum_{i}\frac{R^6}{r^6_{ij}}\left( E_{\medcirc \medcirc} -  E_{\medbullet \medbullet}\right)$, which is to a good approximation constant in the bulk but acts as a pinning field at the open boundaries in an experiment. 

\begin{figure}[t!!]
\centering
\includegraphics[width=\linewidth]{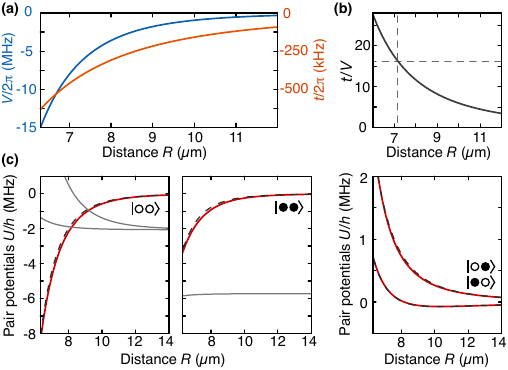}
\caption{\textbf{Realistic experimental parameters.} \textbf{(a)} Extended Hubbard interactions $V/R^6$ (blue) and tunnelings $t/R^3$
(orange) between the proposed Rydberg states at various distances. Similar interactions can be expected to be found for other Rydberg states and atomic species. \textbf{(b)} At a distance close to $R \approx 7.1\,\si{\micro\meter}$ where $V/t \approx 16$, we expect the presence of a supersolid phase. \textbf{(c)} The underlying diagonal Rydberg interactions~\cite{Weber2017} between $\ket{\medbullet \medbullet}$ and $\ket{\medcirc \medcirc}$ (red lines) can be well approximated by $E_{\medbullet \medbullet}(R) \approx C^{\medbullet \medbullet}_6/R^6$ and $E_{\medcirc \medcirc}(R) \approx C^{\medcirc \medcirc}_6/R^6$ (dashed black lines). Gray lines are other pair potentials which remain uncoupled and do not contribute. Interactions between $\ket{\medcirc \medbullet}$ and $\ket{\medbullet \medcirc}$ have diagonal components $\propto 1/R^6$ contributing to $V$, as well as off-diagonal exchange components $\propto 1/R^3$ contributing to $t$, both included in the fit.}
\label{fig-Rydberg}
\end{figure}

For previous quantum simulations focusing on two Rydberg states with identical principal quantum number $n$, tunneling $t$ is typically much stronger than $V$~\cite{Chen2023}.
Realizing a regime where $t < V$ requires the dipole matrix elements between the chosen Rydberg states, which contribute to $t$, to be smaller than the ones contributing to the dispersive van-der-Waals interactions. % Simultaneously, an overall energy scale larger than typical decoherence effects, induced by Rydberg lifetime or motion, is necessary.
Here, we propose Rydberg states with different $n$ where radial dipole matrix elements decay rapidly.
%Because coupling Rydberg states with large $\Delta n$ is experimentally challenging, we also select Rydberg states with identical angular momentum $j = j^\prime = 3/2$ whose matrix elements are naturally smaller.
Additionally to $n$ and the orbital angular momentum $l$ of the Rydberg states, interactions can be fine-tuned by the total electronic angular momentum state $J$, their projection $m_J$, and the magnetic field $B$ applied perpendicular to the atomic plane. 
As one example, we find promising interaction strengths for states $\ket{\medcirc} = \ket{60P_{3/2},m_j = 3/2}$ and $\ket{\medbullet} = \ket{59 D_{3/2},m_j = 3/2}$ and a field amplitude $B = 50\,$G, see Fig.~\ref{fig-Rydberg}.
These states implement a model with attractive, extended Hubbard interactions~$V<0$ and antiferromagnetic tunnelings~$t<0$. Hence the highest energy state (negative temperature regime) realizes the predicted supersolid, which can be prepared adiabatically as we explain below.
We are confident that similar parameters can be found using different parameters and atomic species, opening a wide range of the accessible parameters.

%Similar interaction strengths can be found at different field strengths, which leaves open parameters
%Both, the pinning field and the dipolar corrections are dependent on the setup and choice of Rydberg states in an actual experiment, which are subject to future studies.

%Furthermore, in the calculations of the pair interactions, we find $1/R^6$ corrections from the idealized dipolar~$1/R^3$ dependence of XY interactions at short distances, see Fig.~\ref{fig-Rydberg}(c). 

%\LP{Note that the open boundaries, the pinning fields, and the spin-exchange interactions corrected with a $1/r^6$ term, have not been taken into account into the QMC simulations. Those effects are estimated to be small in the bulk and dependent on the setup and choice of Rydberg states in an actual experiment. }

%%%%%%%%%%%%%%%%%%%%%%%%%%%%%%%%%%
\textbf{State preparation protocol.---}
In Rydberg tweezer experiments, it is customary to prepare an initial product state followed by a ramp protocol in order to prepare a target ground state~\cite{Semeghini2021,Chen2023}. Due to imperfections in the initial state preparation, decoherence effects during the time evolution, e.g., caused by finite Rydberg state lifetime, and rapid spin exchange processes, local observables are often well described by an equilibrium grand canonical statistical ensemble during the later stages of an experiment~\cite{Sbierski2024}. We proceed as follows:
%In order to prepare the supersolid in an experiment, we propose the following adiabatic state preparation protocol, in analogy to previously performed experiments~\cite{Chen2023}: 
(i) All atoms are prepared in the $\ket{\medbullet}$ state. (ii) Subsequently, local light shifts~$|\Delta_j| \gg |t|, |V|$ acting on the~$\ket{\medcirc}$ state are applied to sites~$j\in C_N$, where $C_N$ corresponds to a realization of the ${\rm CDW}_{1/3}$ order~\cite{Scholl2021} with $N$~additional excitations in the system, see Fig.~\ref{fig-overview}(a). Depending on the sign of the light shift, the system has a large overlap with the ground state ($\Delta_j > 0$) or highest excited state ($\Delta_j < 0$).
(iii) In the presence of the strong local light shift, a global microwave pulse applies a $\pi$-rotation~$\hat{U}_{\pi} =\prod_{j \notin C_N}( \hat{a}^\dagger_j +\hat{a}_j)$ for the atoms on site~$j\notin C_N$.
(iv) To adiabatically prepare the ground state (highest excited state), the local light shift~$|\Delta_j|$ is decreased to zero while maintaining an eigenstate for ramp speeds slower than the gap size. We expect that the long-range interactions give rise to large finite system size gaps advantageous for the experiments. Previous experiments in similar systems were mostly limited by infidelities in the initial state preparation and not diabaticity~\cite{Chen2023,Sbierski2024}. To account for non-equilibrium processes, we suggest to wait for a short time to reach thermal equilibrium before preforming measurements.
(v) Lastly, a projective measurement on one internal state is performed. In the above basis, this directly allow us to obtain the $\langle \hat{n}_i \hat{n}_j \rangle$ correlations and static structure factor~$S(\bm{Q})$. An additional global $\pi/2$ rotation around the~$(\hat{a}^\dagger + \hat{a})$-direction prior to the measurement gives access to the correlations of~$\langle \hat{a}^\dagger_i\hat{a}_j + \mathrm{h.c.} \rangle$. 
Therefore, the proposed protocol allows us to directly test our predicted supersolid phase by a comparison to the correlation functions obtained using QMC simulations, see Fig.~\ref{fig:illustr}. 

%%%%%%%%%%%%%%%%%%%%%%%%%%%%%%%%%%
\textbf{Discussion and Outlook.---}
We have introduced a new model combining long-range dipolar tunnelings of hard-core bosons with dominant short-range vdW Hubbard interactions in Rydberg tweezer arrays. On the triangular model and for ferromagnetic tunnelings/repulsive extended Hubbard interactions, we have established the phase diagram using large-scale QMC simulations in the experimentally relevant parameter regime. We have predicted the existence of a robust supersolid phase. From a finite temperature analysis, we have extracted the critical temperature and entropy, which we find to be within reach in current Rydberg analog quantum simulators using concrete calculations of pair states in ${}^{87}\mathrm{Rb}$. Our proposed scheme makes the realization of the long-sought lattice supersolid with hundreds of particles viable.

In addition to the equal time correlation functions, it has been demonstrated~\cite{chen2023spectroscopy} that dynamical experiments give access to the momentum resolved excitation spectrum of spin models~\cite{Knap2013}. The application of this protocol to our model allows to study the emergence of superfluid as well as crystalline phonons in the supersolid phase~\cite{Hertkorn2021}.
Furthermore, the Hamiltonian~\eqref{eq:Hamiltonian-XY-ZZ} can be considered from the perspective of a spin-ordered Mott insulator; hence extensions to three Rydberg levels provide a route to explore supersolids with additional mobile hard-core bosonic vacancies~\cite{Homeier2024}.

Our proposed experimental scheme is highly flexible: The ability to probe either the ground state or the inverted spectrum~\cite{Chen2023} as well as the freedom to engineer both ferromagnetic and antiferromagnetic dipolar tunnelings by choosing atomic~$m_j$ sublevels enables to probe four distinct models. This includes %\LP{longer range spin exchanges of the antiferromagnetic spin supersolids~\cite{Xiang2024} and} 
the recently discovered transverse quantum fluids~\cite{Kuklov2024,Kuklov2024a} with ferromagnetic tunnelings and attractive Hubbard interactions. A sign problem for antiferromagnetic tunnelings prohibits the exploration of large-scale QMC simulation, but frustrated models host potential candidates for quantum spin liquids~\cite{Bintz2024}. Hence their quantum simulation could give valuable insights into their quantum phases of matter. 

%%%%%%%%%%%%%%%%%%%%%%%%%%%%%%%%%%

\textit{Note added.--} During completion of this work we became aware of another proposal to study supersolidity with Rydberg tweezer arrays based on three Rydberg states~\cite{Liu2024}.

\textbf{Acknowledgements.---}
We thank Daniel Barredo, Guillaume Bornet, Cheng Chen, Gabriel Emperauger, Bastien Gély, Fabian Grusdt, Lukas Klein, Thierry Lahaye, Simon Linsel, Mu Qiao, Ana Maria Rey and Norman Yao for fruitful discussions. This research was funded by the Deutsche Forschungsgemeinschaft (DFG, German Research Foundation) under Germany's Excellence Strategy -- EXC-2111 -- 390814868. L.H. has received funding from the European Research Council (ERC) under the European Union’s Horizon 2020 research and innovation programm (Grant Agreement no 948141) — ERC Starting Grant SimUcQuam, and acknowledges support from the Studienstiftung des deutschen Volkes. This work was also partly supported by the Simons Collaboration on Ultra-Quantum Matter, which is a grant from the Simons Foundation (651440). S.H. acknowledges funding through the Harvard Quantum Initiative Postdoctoral Fellowship in Quantum Science and Engineering. S.G. acknowledges funding by Structures (Deutsche Forschungsgemeinschaft (DFG, German Research Foundation) under Germany’s Excellence Strategy EXC2181/1-390900948). A.B. acknowledges funding by the Agence Nationale de la Recherche (ANR-22-PETQ-0004 France 2030, project QuBitAF), Horizon Europe programme HORIZON-CL4-2022-QUANTUM-02- SGA via the project 101113690 (PASQuanS2.1), and the European Research Council (Advanced grant No. 101018511- ATARAXIA).

%% references
%\section*{References}
%\bibliographystyle{apsrev4-1}

%% supplement
\pagebreak
\clearpage
\newpage
\appendix
%\onecolumngrid
\widetext
\begin{center} 
\textbf{\large Supplemental Materials: Supersolidity in Rydberg tweezer arrays}
\end{center}
%%%%%%%%%% Merge with supplemental materials %%%%%%%%%%
%%%%%%%%%% Prefix a "S" to all equations, figures, tables and reset the counter %%%%%%%%%%
\setcounter{equation}{0}
\setcounter{figure}{0}
\setcounter{table}{0}
\setcounter{page}{1}
\makeatletter
\renewcommand{\theequation}{S\arabic{equation}}
\renewcommand{\thefigure}{S\arabic{figure}}

\section{Selection of Rydberg states}
In this manuscript, we predict the presence of supersolid phases in atomic arrays where all atoms are excited to their Rydberg states. We focus on systems of two Rydberg states with opposite parity, where the orbital angular momentum $l$ between both states differs by one, i.e., $\Delta l = 1$.
Here, resonant dipole-dipole interactions between atom pairs are typically significantly stronger than dispersive van-der-Waals interactions, which emerge from second-order dipole-dipole interactions to off-resonant pair potentials. 
We propose to use two Rydberg states with different principal quantum numbers $\Delta n \neq 0$ where the dipole matrix element between both Rydberg states decreases drastically.
This enables us to access the opposite regime where van-der-Waals interactions dominate and supersolids are expected, as we confirm using large-scale QMC simulations. 

We studied various Rydberg states $\ket{nS_{1/2}}$, $\ket{nP_{J}}$ and $\ket{nD_{J}}$ of $^{87}$Rb at different principle quantum numbers $n$ and $n'$. 
For Rydberg atom pairs $\ket{nS_{1/2}}$ and $\ket{n'P_{J}}$, for typical principal quantum numbers, resonant dipole-dipole interactions decrease too quickly with $\Delta n$.
As a result, $t/V$ is either too large, such that we do not expect the presence of a supersolid phase, or too small, such that it is hard to observe experimentally.
For states $\ket{nD_{J}}$ and $\ket{n'P_{J}}$, we predict the interesting parameter regime if $n = n' - 1 $. 
For the relevant principal quantum numbers, both Rydberg states are energetically less than $10\,$GHz separated, enabling an efficient coupling using state-of-the art microwave technology.

We further fine-tune the interactions by the magnetic quantum numbers $m_J$ as well as the magnetic field $B$. 
We chose the magnetic field to point perpendicular to the atomic plane such that interactions between atoms in the atomic plane are independent of the orientation of the interacting atom pairs. 
The dipole-dipole interactions furthermore depend on the magnitude of the magnetic field $B$ because it mixes the fine-structure levels of both Rydberg states, which affects their dipole matrix elements.
An additional constraint is the relative sign of $t$ and $V$, which depends on $m_J$.
We only expect the system to support a supersolid phase if $t/V > 0$.
Finally, we converged to states $\ket{\medcirc } = \ket{60P_{3/2},m_j = 3/2}$ and $\ket{\medbullet } = \ket{59 D_{3/2},m_j = 3/2}$ and a field amplitude $B = 50\,$G.
These states have the additional advantage that the van-der-Waals interactions between D-state atom pairs is relatively weak.
This enables an efficient excitation of the atomic array into the state $\ket{\medbullet } $, which is an essential part of the proposed state preparation. 

In Fig.\,\ref{fig-Rydberg} in the main text, it has been discussed that interactions between Rydberg pairs $\ket{\medcirc \medbullet}$ and $\ket{\medbullet \medcirc }$ contain a resonant off-diagonal term $\propto 1/R^3$, which induces dipolar exchange and mixes both terms, as well as a diagonal contribution $1/R^6$. 
At short distances, we expect modifications from additional contributions (such as off-diagonal exchange interactions $\propto 1/R^6$). These terms were small for our specific Rydberg pairs but are generally not zero. 

\section{Phase Diagram}

\begin{figure}[h]
\centering
\includegraphics[width=0.6\linewidth]{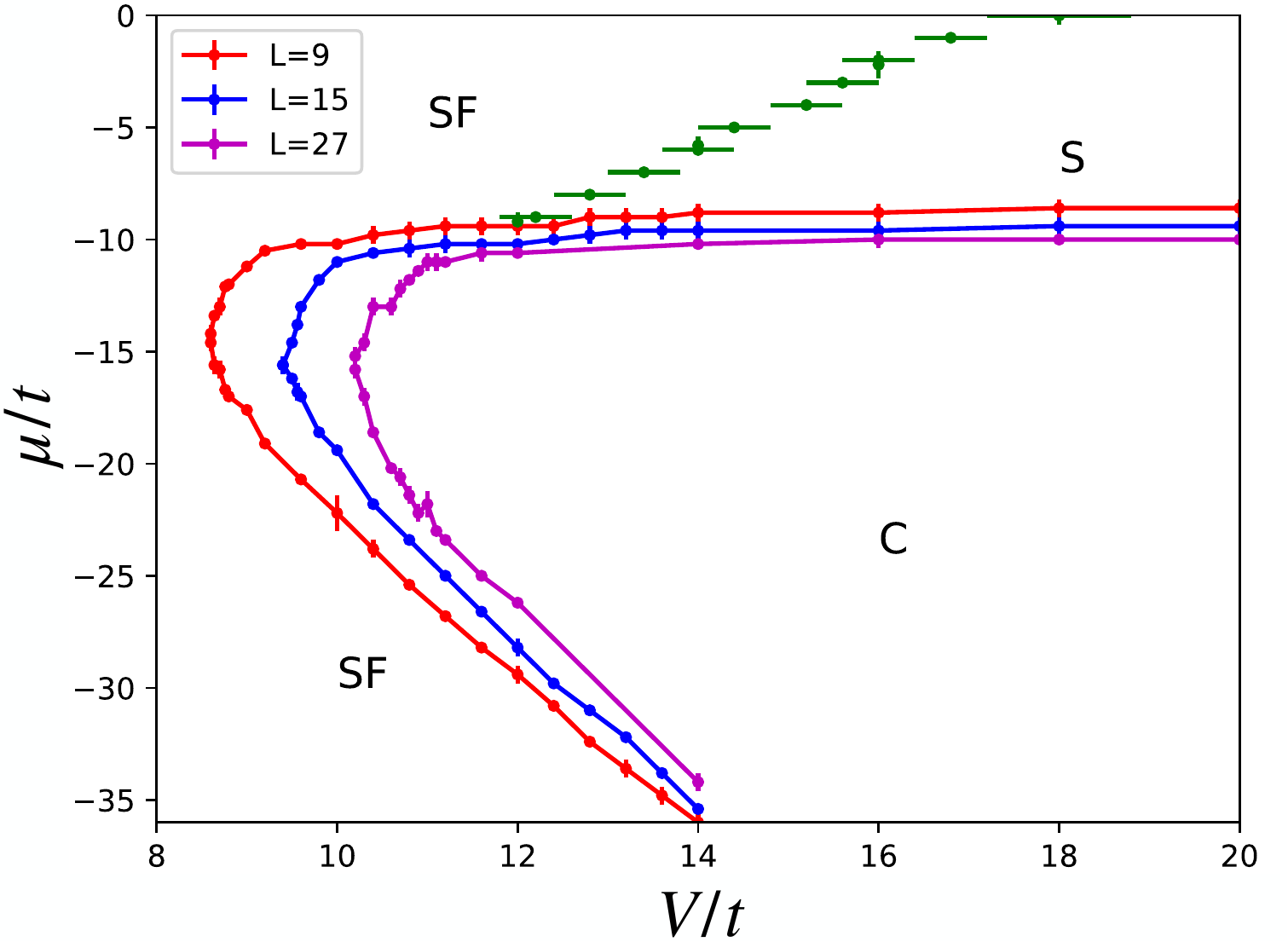}
\caption{\textbf{Phase diagram for $\beta t = 5$ in the $(V/t, \mu/t)$-plane.} The transition (shown in green) between the superfluid (SF) and supersolid (S) phase is second order and based on finite size scaling for scans along $\mu/t$ (vertical error bars) and $V/t$ (horizontal error bars).  The insulating charge density wave phase (C) with filling $1/3$ is shown for various system sizes $L$ in order to show the strong finite size effects near the tip of the CDW lobe. Transitions between CDW and supersolid are second order. Lines are a guide to the eye. } 
\label{fig_CDW}
\end{figure}

In order to appreciate finite size effects we show the phase diagram of the model in Fig.~\ref{fig_CDW} obtained with quantum Monte Carlo simulations. We took an inverse temperature $\beta t = 5$ which is close to the ground state. The transitions between the superfluid and supersolid phase are of second order and determined from a finite size analysis, i.e., extrapolated to the thermodynamic limit. The insulating, crystalline CDW phase for different system sizes was determined from a requirement that the superfluid stiffness~$\rho_S < 0.1$. The superfluid stiffness is computed in quantum Monte Carlo simulations from  the fluctuations of the winding number as $\chi_{\rm s} = \left< W^2 \right> / (2 \beta)$~\cite{Pollock1987}. The transition between the insulating crystalline phase and the superfluid phase is first order. It is particularly strong away from the tip of the lobe. The shrinking of the Mott lobe, in particular in the vicinity of the tip of the lobe, is remarkably strong. For instance, a scan in $\mu/t$ at fixed $V/t=10$ shows a charge density wave lobe for $L=9$ and $L=15$ but (just) not for $L=27$. Such finite size effects will affect a future experiment, which will have comparable system sizes as the ones shown here. These finite size effects originate from the kinetic, dipolar term in the Hamiltonian, which is parameterically stronger than the potential term, and thus grows stronger in importance with increasing $L$ at low enough temperature.

\section{Finite size scaling analysis at finite temperature}

\begin{figure}[h]
\centering
\includegraphics[width=0.6\linewidth]{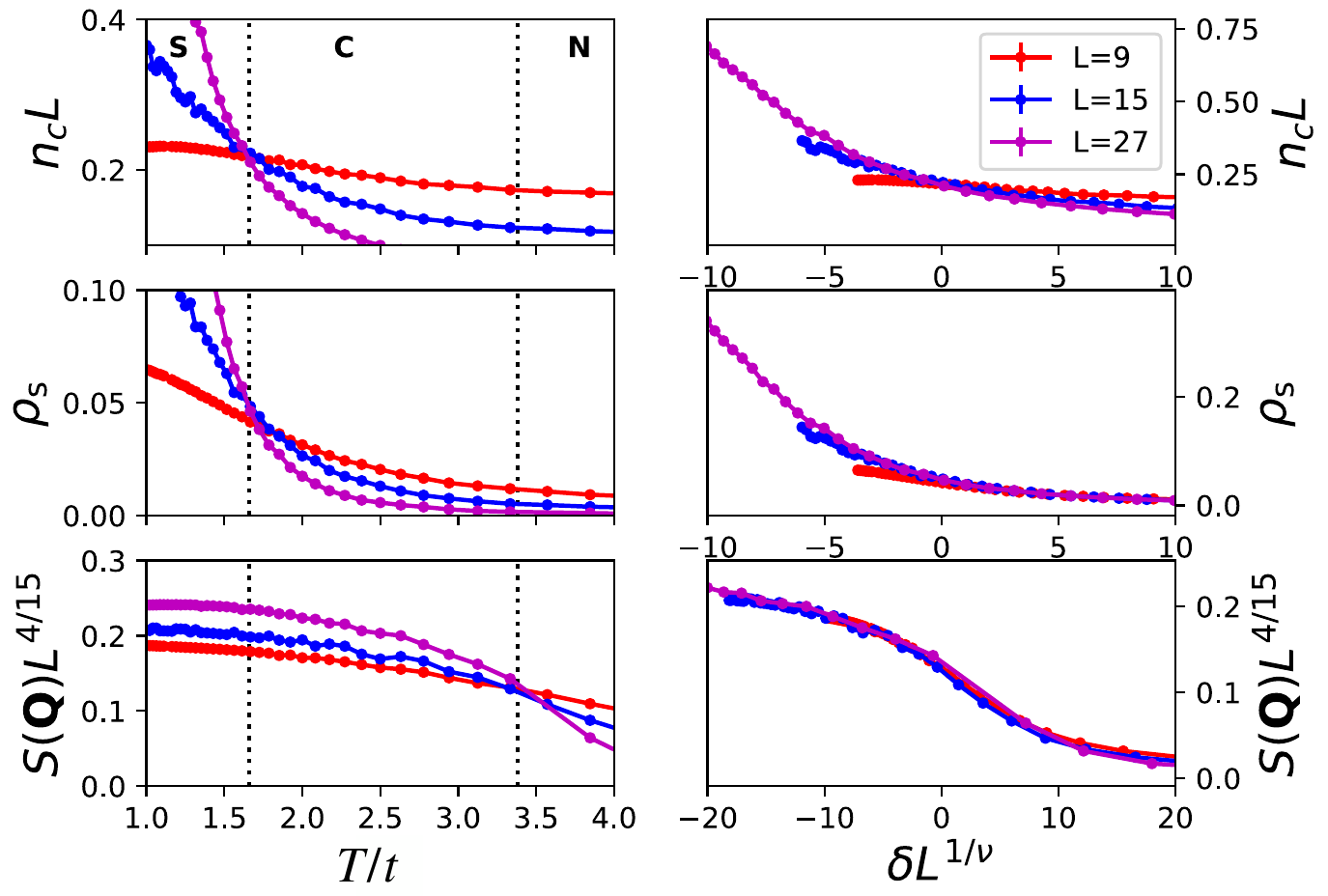}
\caption{\textbf{Finite size scaling analysis of the finite temperature transitions} for $V/t=20$, $\mu/t= -9$, identifying the normal (N), crystalline (C), and supersolid (S) phases. The upper two rows show consistent information for the $U(1)$ transition, based on the condensate density and the superfluid stiffness, respectively. The lower row indicates the Potts transition based on the static structure factor. On the left, the intersection point determines the critical temperature. On the right, the collapse of the curves for different system sizes confirms the critical exponents. Note that the value of $\nu$ on the $x-$label differs for the $U(1)$ and the Potts transitions.}
\label{fig_temperature3}
\end{figure}

For the $U(1)$ transitions at finite temperature, we expect the critical exponents $\nu = \eta = 1$ as has been shown in Ref.~\cite{Sbierski2024}. The scale invariant quantities are the superfluid stiffness $\rho_s$ and the condensate density scaled wiht the linear system size, $n_c L$. The crystalline (CDW) transition at finite temperature belongs to the 2D 3-state Potts model with known critical exponents $\beta=1/9$ and $\nu = 5/6$. The finite size scaling analysis is then performed for the rescaled static structure factor, $S(\bm{Q})L^{2\beta/\nu}=S(\bm{Q})L^{4/15}$, at the ordering wavevector $\bm{Q}$.
In Fig.~\ref{fig_temperature3} we confirm that these exponents are compatible with the quantum Monte Carlo data.

\begin{figure}[h]
\centering
\includegraphics[width=0.7\linewidth]{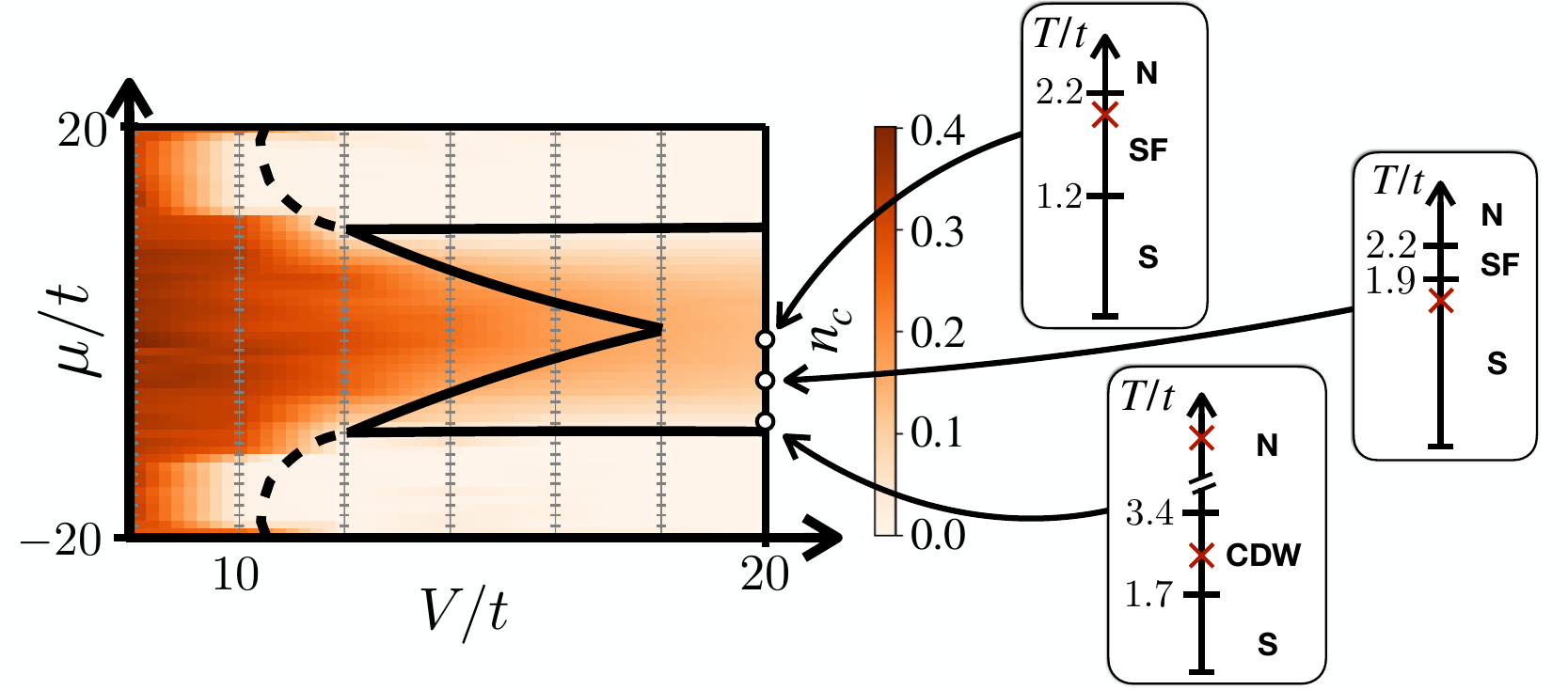}
\caption{\textbf{Finite temperature transitions.} We consider the order of finite temperature transitions at $V/t=20$ and for $\mu/t=-1,-5,-9$. Close to the charge density wave order, we find that first the translational symmetry and then the $U(1)$~symmetry is broken (right bottom). Vice versa, close to half filling first the $U(1)$~symmetry and then the translational symmetry is broken (right center and top). The red crosses indicate the parameters used in Fig.~\ref{fig:illustr}.} 
\label{fig_T-cuts}
\end{figure}

\section{Order of finite temperature transitions}

In our numerical Monte Carlo simulation, we address the finite temperature phase diagram of Hamiltonian~\eqref{eq:Hamiltonian-XY-ZZ}. The supersolid phase is characterized by the spontaneous symmetry breaking of the translation lattice symmetry and the continuous $U(1)$~symmetry. Approaching the supersolid from high temperatures, we find that (i) the two symmetry breaking transitions typically occur at different temperatures and (ii) the ordering of the transition depends on the chemical potential~$\mu$.

In Fig.~\ref{fig_T-cuts}, we fix the extended Hubbard interaction~$V/t=20$ (corresponding to a fixed distance between Rydberg atoms) and consider three values of the chemical potential~$\mu/t=-1,-5,-9$, i.e. the parameters used in the main text in Fig.~\ref{fig:illustr}. In the vicinity of the CDW phase ($\mu/t=-9$), the system first exhibits a structure factor at higher temperature before the additional defects condense into the supersolid at low temperature. In contrast, closer to half-filling at~$\mu/t=-5$ the transition temperature of the crystalline phase decreases while simultaneously the superfluid's critical Tc increases, such that we find a reversed ordering of the transition: The system establishes a finite condensate fraction before ordering to a supersolid as temperature is lowered. 
We explain this behaviour by the increasing number of defects closer to half filling, which leads to a larger superfluid density causing a higher transition temperature for the $U(1)$~symmetry breaking. In contrast, the critical temperature for the Potts transition is decreasing as the system is doped away from the $1/3$-CDW, such that we find the highest critical temperatures of the supersolid phase between CDW ordered ground state and half filling.

\end{document}